\documentclass[a4paper]{jpconf}
\usepackage{graphicx}

\newcommand{\be}{\begin{equation}}
\newcommand{\ee}{\end{equation}}
\newcommand{\bea}{\begin{eqnarray}}
\newcommand{\eea}{\end{eqnarray}}

\begin{document}
\title{Dilepton production from SIS to LHC energies}

\author{E L Bratkovskaya$^{1,2}$,  O Linnyk$^3$, V P Konchakovski$^3$, W
Cassing$^3$,  \\
V Ozvenchuk$^{2}$, J Manninen$^{1,2}$ and C M Ko$^4$ }

\address{$^1$ Institute for Theoretical Physics, University of
Frankfurt, Frankfurt, Germany}

\address{$^2$ Frankfurt Institute for Advanced Studies, %
 60438 Frankfurt am Main}

\address{$^3$ Institute for Theoretical Physics, University of Giessen,
Giessen, Germany}

\address{$^4$ Texas A\& M Universtity, Texas, U.S.A.}

\ead{Elena.Bratkovskaya@th.physik.uni-frankfurt.de}

\begin{abstract}
We study $e^+e^-$ pair production in proton-proton and in
nucleus-nucleus collisions from SIS to LHC energies within the
parton-hadron-string dynamics (PHSD) approach which incorporates
explicit partonic degrees-of-freedom in terms of strongly interacting
quasiparticles (quarks and gluons) in line with an equation-of-state
from lattice QCD as well as the dynamical hadronization and hadronic
collision dynamics in the final reaction phase. We find a visible
in-medium effect in the low mass dilepton sector from dynamical
vector-meson spectral functions from SIS to SPS energies whereas at
RHIC and LHC energies such medium effects become more moderate.  In the
intermediate mass regime from 1.1 to 3 GeV pronounced traces of the
partonic degrees of freedom are found at SPS and RHIC energies which
superseed the hadronic (multi-meson) channels as well as the correlated
and uncorrelated semi-leptonic $D$-meson decays. The dilepton
production from the strongly interacting quark gluon plasma (sQGP)
becomes already visible at top SPS energies and more pronounced at RHIC
and LHC energies.
\end{abstract}

\section{Introduction}
Dileptons, i.e. correlated  electron and positron pairs, are one of the
key observables in ultra-relativistic nuclear collisions experiments
since dileptons are emitted during the whole collision evolution and
thus one may probe various aspects at different stages of a
relativistic nuclear collision by measuring differential dilepton
spectra. Another important feature is that the produced leptons
interact only electromagnetically and thus  interact only very weakly
with the strongly interacting partonic or hadronic medium created in
the collisions. In another words, e.g. an initial state Drell-Yan-pair
from the early stages of the collision is expected to survive the
 subsequent evolution of the fireball.

An important dilepton observable is the invariant mass spectrum. The
mass spectrum can be roughly divided in 3 different regions, in each of
which different physics dominates the radiation. In the low mass region
($M_{\e^+\e^-}<$ 1 GeV) the radiation is dominated by the decays of
light mesons (consisting of $u$, $d$ and $s$ (anti)quarks) and
especially gives information about in-medium properties of the $\rho^0$
meson. In the intermediate mass region (1.1 GeV $<M_{\e^+\e^-}<$ 3 GeV)
the dominant hadronic contribution to the invariant mass spectrum is obtained
from the decays of open charm mesons, while above the $J/\psi$ peak,
first open beauty decays and later on initial state Drell-Yan radiation
are expected to dominate the dilepton spectrum. On top of the
previously mentioned sources, especially in the intermediate mass
region, also radiation from the strongly interacting Quark-Gluon-Plasma
(sQGP) can give a significant signal~\cite{Shuryak:1978ij} as well as
some other more exotic sources like simultaneous interactions of four
pions~\cite{Song:1994zs,Li:1998ma},\cite{vanHees:2006ng,vanHees:2007th}.
These partonic and hadronic channels have been studied in detail in
Refs. \cite{Linnyk:2011vx,Linnyk:2011hz} at the top
Super-Proton-Synchrotron (SPS) and Relativistic-Heavy-Ion-Collider
(RHIC) energies and it has been found that the partonic channels
clearly dominate over multi-pion sources in the intermediate dilepton
mass regime.
This contribution aims to summarize the perspectives of dilepton
measurements from low SIS to high LHC energies
based on the Parton-Hadron-String Dynamics (PHSD) transport
model~\cite{PHSD}.

\section{The PHSD approach}

The dynamics of partons, hadrons and strings in relativistic
nucleus-nucleus collisions is analyzed here within the
Parton-Hadron-String Dynamics approach~\cite{PHSD}. In this
transport approach the partonic dynamics is based on Kadanoff-Baym
equations for Green functions with self-energies from the Dynamical
QuasiParticle Model (DQPM) \cite{Cassing07} which
describes QCD properties in terms of 'resummed' single-particle
Green functions. In Ref.~\cite{BCKL11}, the actual  three DQPM
parameters for the temperature-dependent effective coupling were
fitted to the recent lattice QCD results of Ref.~\cite{aori10}.
The latter lead to a critical temperature $T_c \approx$ 160 MeV
which corresponds to a critical energy density of $\epsilon_c
\approx$ 0.5 GeV/fm$^3$. In PHSD the parton spectral functions
$\rho_j$ ($j=q, {\bar q}, g$) are no longer $\delta-$ functions in
the invariant mass squared as in conventional cascade or transport
models but depend on the parton mass and width parameters which
were fixed by fitting the lattice QCD results from
Ref.~\cite{aori10}. We recall that the DQPM allows one to extract a potential
energy density $V_p$ from the space-like part of the energy-momentum
tensor as a function of the scalar parton density $\rho_s$.
Derivatives of $V_p$ w.r.t. $\rho_s$ then define a scalar mean-field
potential $U_s(\rho_s)$ which enters into the equation of motion for the
dynamical partonic quasiparticles. Furthermore, a two-body
interaction strength can be extracted from the DQPM as well from the
quasiparticle width in line with Ref.~\cite{Andre}. The transition
from partonic to hadronic d.o.f. (and vice versa)  is
described by covariant transition rates for the fusion of
quark-antiquark pairs or three quarks (antiquarks), respectively,
obeying flavor current-conservation, color neutrality as well as
energy-momentum conservation~\cite{PHSD,BCKL11}. Since the dynamical
quarks and antiquarks become very massive close to the phase
transition, the formed resonant 'prehadronic' color-dipole states
($q\bar{q}$ or $qqq$) are of high invariant mass, too, and
sequentially decay to the groundstate meson and baryon octets
increasing the total entropy.

On the hadronic side PHSD includes explicitly the  baryon octet and
decouplet, the $0^-$- and $1^-$-meson nonets as well as selected
higher resonances as in the Hadron-String-Dynamics (HSD) approach
 \cite{Ehehalt,HSD}. Hadrons of higher
masses ($>$ 1.5 GeV in case of baryons and $>$ 1.3 GeV for
mesons) are treated as 'strings' (color-dipoles) that  decay to the
known (low-mass) hadrons, according to the JETSET algorithm
\cite{JETSET}. Note that PHSD and HSD  merge at low energy density, in
particular below the critical energy density $\epsilon_c \approx 0.5 $
GeV/fm$^{3}$.
For more detailed descriptions of PHSD and its ingredients we refer the
reader to Refs. \cite{Cassing07,BCKL11,Cassing06,Cas09}.

The PHSD approach was applied to nucleus-nucleus collisions
from $s_{NN}^{1/2} \sim$ 5 to 200 GeV in Refs.~\cite{PHSD,BCKL11} in
order to explore the space-time regions of partonic matter. It was
found that even central collisions at the top-SPS energy of
$\sqrt{s_{NN}}=$17.3 GeV show a large fraction of nonpartonic, {\it
i.e.} hadronic or string-like matter, which can be viewed as a
hadronic corona. This finding implies that neither hadronic nor only
partonic 'models' can be employed to extract physical conclusions in
comparing model results with data.

\section{Results for dilepton spectra in comparison to experimental data}

We directly continue with the results from PHSD in comparison with the
available experimental data on dilepton production from SIS to RHIC
energies.

\subsection{SIS energies}
\begin{figure}[!t]
\centerline{\includegraphics[width=6.cm]{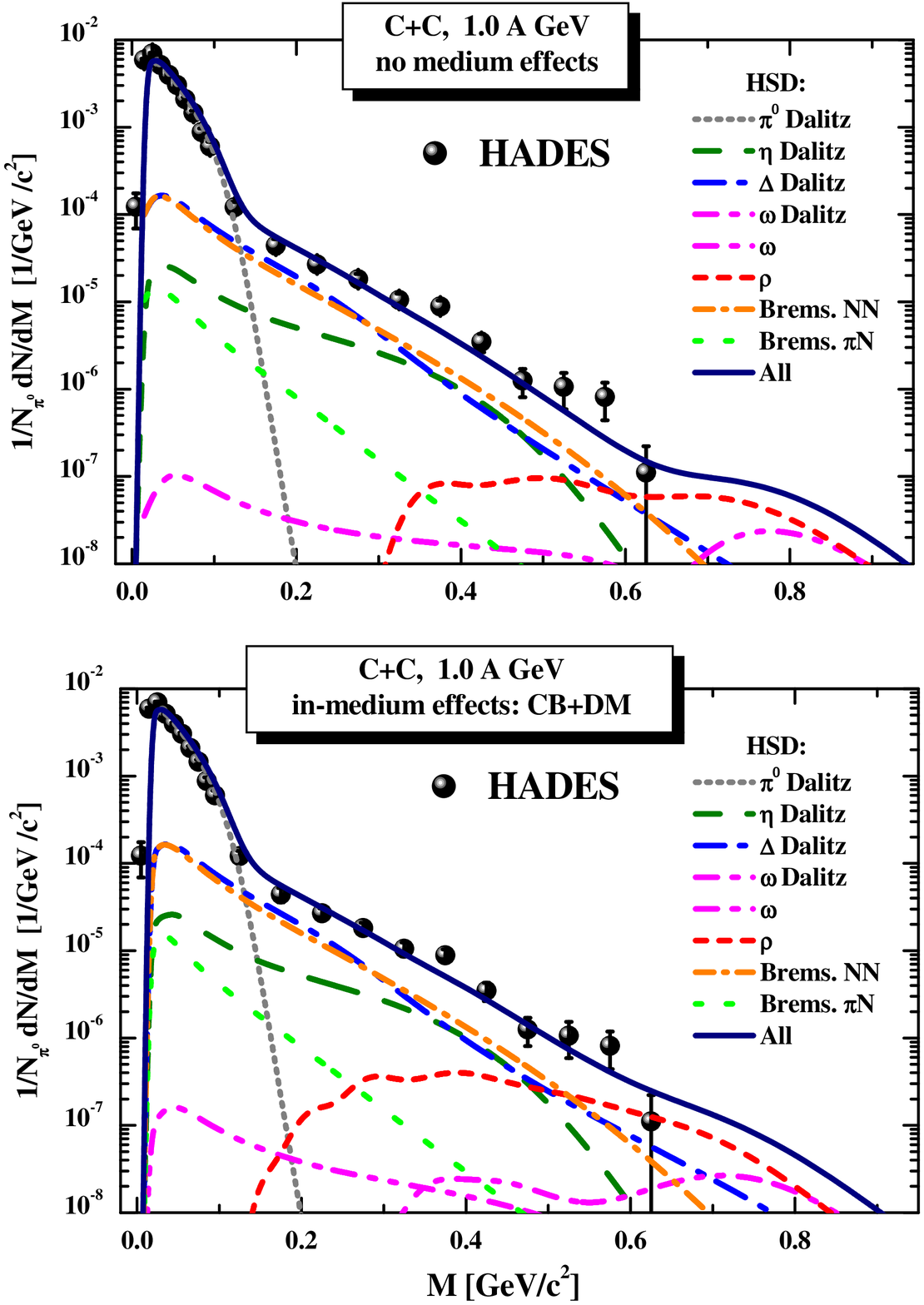}
\includegraphics[width=6.1cm]{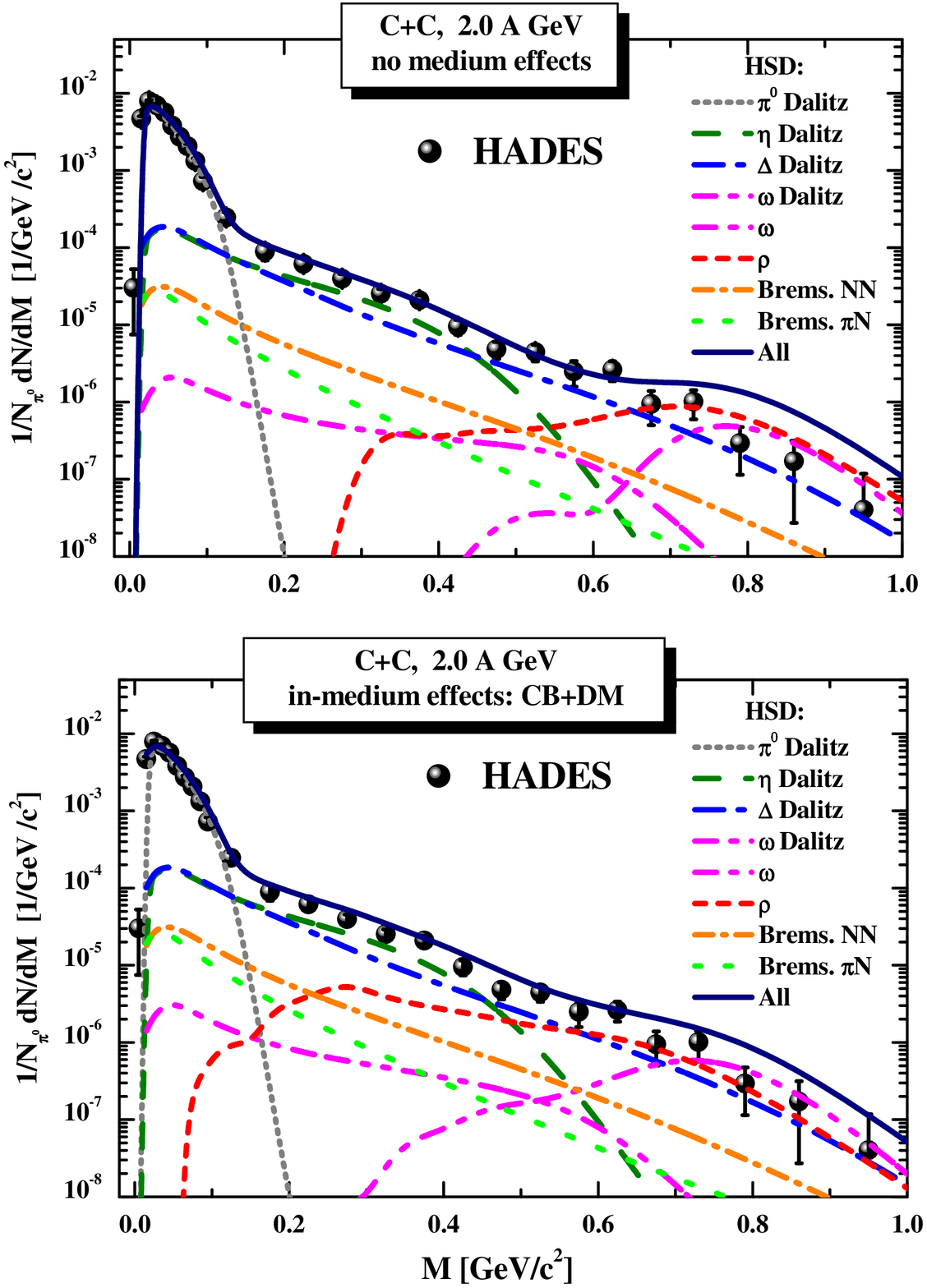}}
\caption{Results of the HSD transport calculation  for the mass
differential dilepton spectra - divided by the average number of
$\pi^0$'s - in case of $^{12}C+^{12}C$ at 1.0 A GeV (left) and
2.0 A GeV (right)  in
comparison to the HADES data \cite{HADES06,HADES07}. The upper part shows the case
of 'free' vector-meson spectral functions while the lower part gives
the result for the 'dropping mass + collisional broadening'
scenario. In both scenarios the HADES acceptance filter and mass
resolution have been incorporated. The different color lines
display individual channels in the transport calculation (see
legend). } \label{Fig13}
\end{figure}
The dileptons produced in low energy heavy-ion collisions
have been measured  first  by the DLS Collaboration at Berkeley
\cite{Matis:1994tg,Wilson:1997sr,Wilson:1993mp,Porter:1997rc}.
The observed dilepton yield  \cite{Porter:1997rc} in the mass range
from 0.2 to 0.5 GeV in  C+C and Ca+Ca collisions at 1 A GeV
was about of five times higher than the calculations by
different transport models
using the 'conventional' dilepton sources as bremsstrahlung, $\pi^0,
\eta, \omega$ and $\Delta$ Dalitz decays and direct vector mesons
($\rho, \omega, \phi$) decays
\cite{Wolf:1992gg,Bratkovskaya:1996bv,Xiong:1990bg}.
Even after including the different in-medium scenarios as collisional broadening and dropping mass
for the $\rho$-meson spectral function did not solve the 'DLS puzzle'
\cite{Ernst:1997yy,BratRapp98,BratKo99,Fuchs:2005zga}.

The recent experimental data from the HADES Collaboration at GSI
\cite{HADES07,Pachmayer:2008yn,Sudol:2008zz,Agakishiev:2009yf,Lapidus:2009aa,Agakishiev:2011vf},
however, confirmed the measurement of the DLS Collaboration
for  C+C at 1.0 A GeV  \cite{Pachmayer:2008yn} as well as for the
elementary reactions \cite{HADES_pp22}.
 In the mean time also
the theoretical transport approaches as well as effective models for
the elementary $NN$ reactions have been further developed.

A possible solution of the 'DLS puzzle' from the theoretical side
has been suggested in Ref.  \cite{Bratkovskaya:2007jk} by
incorporating stronger $pn$ and $pp$ bremsstrahlung
contributions in line with the  updated One-Boson-Exchange (OBE) model
calculations from Ref. \cite{Kaptari:2005qz}.
As shown in Ref. \cite{Bratkovskaya:2007jk}  the
results of the HSD model (off-shell Hadron-String-Dynamics
(HSD) transport approach)  with  'enhanced' bremsstrahlung cross
sections agree very well  with the HADES  data for C+C
at 1 and 2 A GeV as well as with the DLS data for C + C and
Ca + Ca at 1 A GeV,  especially when including
a collisional broadening in the vector-meson spectral functions.
A similar finding has been obtained by other independent transport groups
-- IQMD \cite{Thomere:2007cj} and Rossendorf BUU \cite{Barz:2009yz}.

Fig. \ref{Fig13} shows a comparison of
the HSD results to the HADES mass-differential dilepton spectra for C +
C at 1.0 A GeV (left) and 2.0 A GeV (right)  \cite{HADES06,HADES07} for
 the  'free' (upper part) and the in-medium scenario (lower part).  At
the higher bombarding energy the $\eta$ Dalitz decay provides the
dominant contribution in the mass region from 0.2 to 0.5 GeV followed
by $\Delta$ Dalitz decays and the combined bremsstrahlung channels. The
mass region around 0.75 GeV is overestimated in the 'free' scenario,
whereas including in-medium spectral functions for the vector mesons
the description of the data is improved  due to shifting of the strength
from the vector-meson pole mass regime to lower invariant mass.
However, the in-medium effects for the light C + C system are only very
moderate.  In order to observe a strong broadening of the $\rho$-meson
spectral function one has to investigate a larger size system such as
$Au+Au$. A corresponding measurement has been performed recently by the
HADES Collaboration and the upcoming data will provide more accurate
information on the in-medium effects.

\begin{figure}[t]
\centerline{\includegraphics[width=10cm]{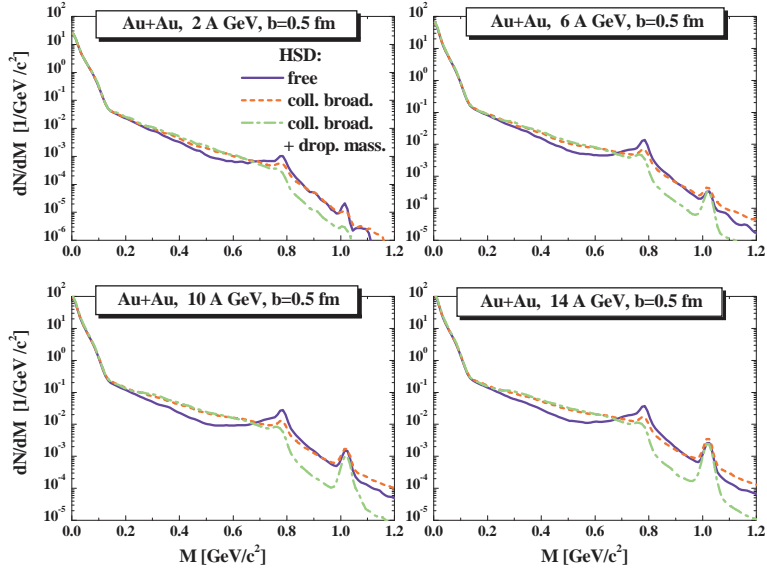}}
\caption{Results of the HSD transport calculation  for the mass
differential dilepton spectra for  central Au+Au collisons
from 2 to 14 A GeV calculated for different in-medium scenarios -
collisional broadening and combined scenario
(dropping mass + collisional broadening). }
\label{Dil_Au}
\end{figure}
In Fig. \ref{Dil_Au} we show the HSD predictions for the dilepton
yields from central Au + Au collisions calculated for different
energies from 2 to 14 A GeV applying the different in-medium
scenarios: collisional broadening and combined approach
(dropping mass + collisional broadening). One can see that both
scenarios lead to an enhancement of the dilepton yield in the mass
region $M =0.3-0.8$ GeV  by a factor of about 2. The largest in-medium
effect is, however, attributed to the reduction of the dilepton yield
between the $\omega$ and $\phi$ peaks due to the downward shift of the
poles of the $\rho$ and $\omega$ spectral functions.  However, the latest
scenario is not consistent with existing experimental data at higher
energies, so one has to rely most likely on the relatively modest
in-medium effects due to  collisional broadening.

\subsection{SPS energies}
We step up in energy and compare our model results with
experimental data for dileptons from
In+In collisions at 160 A GeV measured by the NA60 Collaboration.

\begin{figure}
\begin{minipage} [c] {8.5cm}
\includegraphics[width=0.85\textwidth]{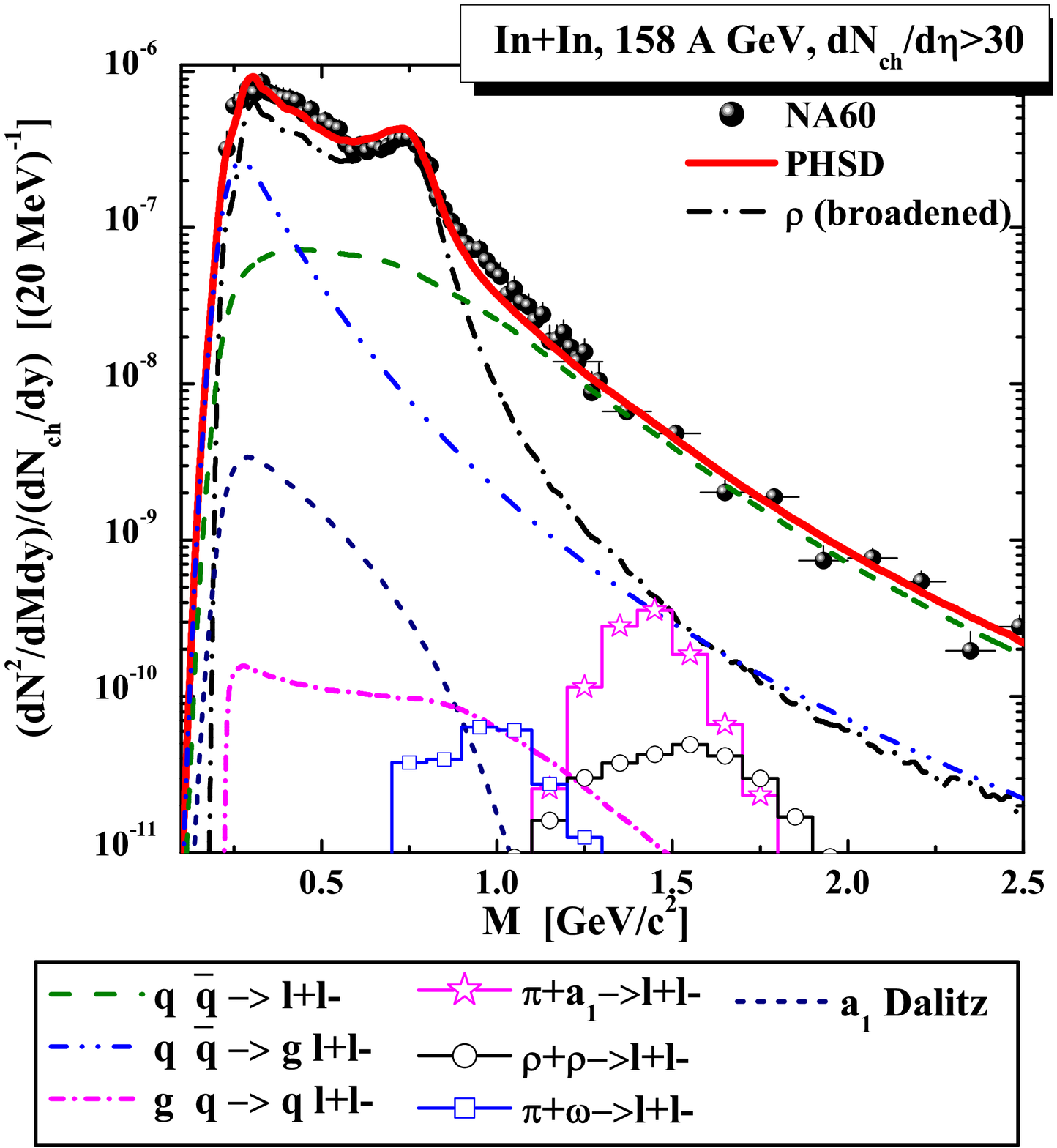}
\end{minipage}
\begin{minipage} [l] {7.cm}
\phantom{a}\vspace*{-10mm}
 \includegraphics[width=1\textwidth]{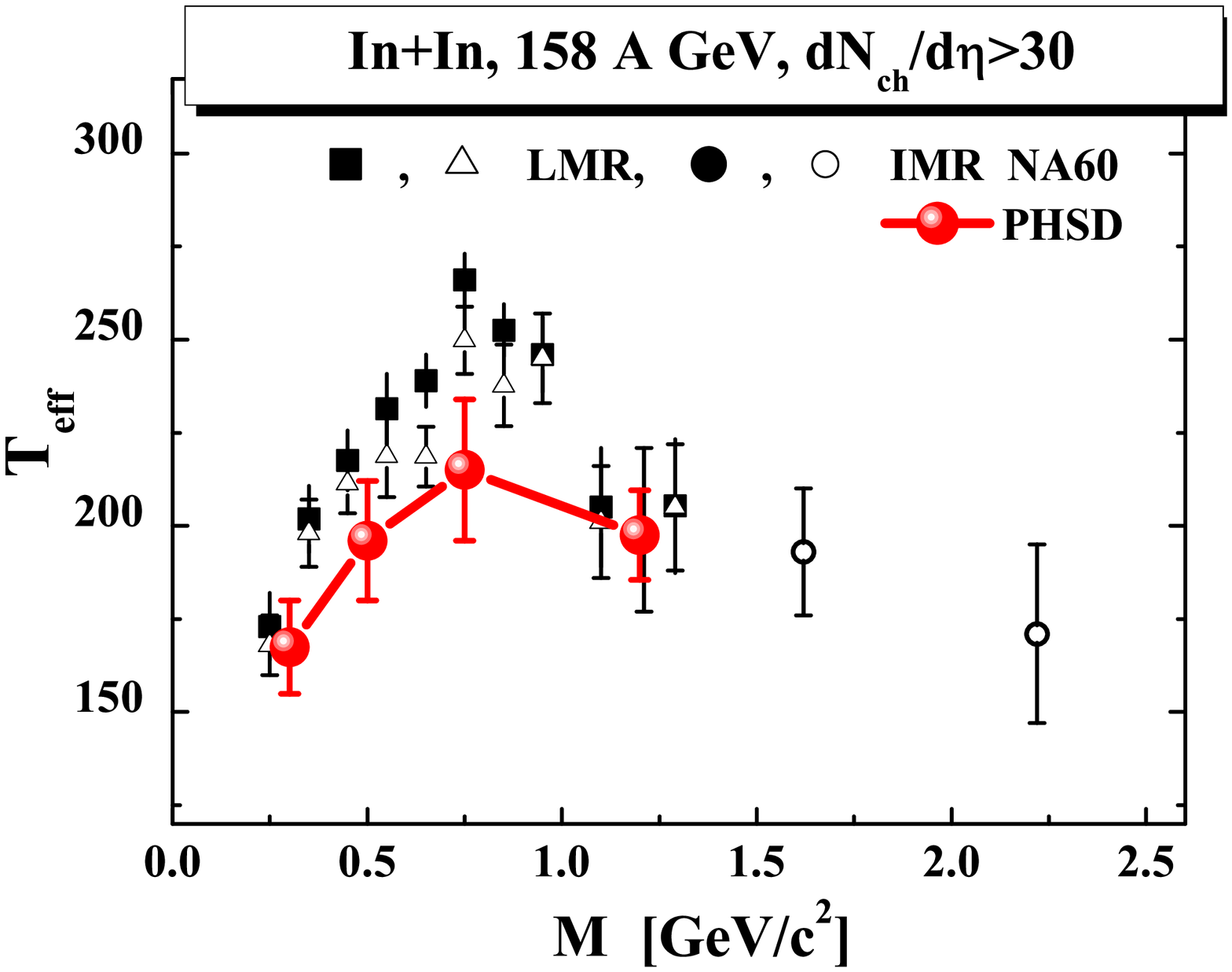}
\end{minipage}
\caption{Left: Acceptance corrected mass spectra of excess
dimuons from In+In at 158~A GeV integrated over $p_T$ in
$0.2<p_T<2.4$~GeV from PHSD compared to the data of
NA60~\cite{Arnaldi:2008er}. The dash-dotted line shows the dilepton
yield from the in-medium $\rho$ with a broadened spectral function,
the dashed line presents the yield from the $q+\bar q$ annihilation,
the dash-dot-dot line gives the contribution of the gluon
Bremsstrahlung process ($q\bar q\to g l^+l^-$), while the solid line
is the sum of all contributions. For the description of the other
lines, which correspond to the non-dominant channels, we refer to
the figure legend.
Right: The inverse slope parameter $T_{eff}$ of the
dimuon yield from In+In at 158 A GeV as a function of the
dimuon invariant mass in PHSD compared to the data of the NA60
Collaboration~\protect{\cite{NA60,Arnaldi:2008er}}.}
\label{NA60_AC}
\end{figure}
In Fig.~\ref{NA60_AC} we present PHSD results for the dilepton
excess over the known hadronic sources as produced in In+In
reactions at 158~A GeV compared to the acceptance corrected
data. We find here that the spectrum at invariant masses in the
vicinity of the $\rho$ peak is well reproduced by the $\rho$ meson
yield, if a broadening of the meson spectral function in the medium
is assumed, while the partonic sources account for the yield at high
masses. Our analysis shows that the contributions of the `$4\pi$'
processes (shown by the lines with symbols), first noted
by the authors of Ref.~\cite{Song:1994zs}, are very much supressed.

One concludes from Fig.~\ref{NA60_AC} that the measured spectrum for
$M>1$~GeV is dominated by the {\it partonic} sources. Indeed, the
domination of the radiation from {the} QGP over the hadronic sources
in PHSD is related to a rather long -- of the order or 3 fm/c --
evolution in the partonic phase (in co-existence with the space-time
separated hadronic phase) on one hand (cf. Fig.~10 of
Ref.~\cite{CasBrat}) and the rather high initial energy densities
created in the collision on the other hand (cf. Fig.~6 of
Ref.~\cite{Linnyk:2008hp}).
In addition, we find from Fig.~\ref{NA60_AC} that in PHSD the
partonic sources also have a considerable contribution to the dilepton
yield for $M<0.6$~GeV. The yield from the two-to-two process $q+\bar
q\to g+l^+l^-$ is especially important close to the threshold ($
\approx 0.211$ GeV). This conclusion from the microscopic
calculation is in qualitative agreement with the findings of an
early (more schematic) investigation in Ref.\cite{Alam:2009da}.

The comparison of the mass dependence of the slope parameter evolution
in PHSD and the data is shown explicitly in the right part of
Fig.~\ref{NA60_AC}.  Including partonic dilepton sources allows us to
reproduce in PHSD the $m_T$-spectra as well as the finding of the NA60
Collaboration~\cite{Arnaldi:2008er,NA60} that the effective temperature
of the dileptons (slope parameters) in the intermediate mass range is
lower than that of the dileptons in the mass bin $0.6<M<1$~GeV, which
is dominated by hadronic sources (cf.  Fig.~\ref{NA60_AC}, right).  The
softening of the transverse mass spectrum with growing invariant mass
implies that the partonic channels occur dominantly before the
collective radial flow has developed. Also, the fact that the slope in
the lowest mass bin and the highest one are approximately equal -- both
in the data and in PHSD -- can be traced back to the two windows of the
mass spectrum that in our picture are influenced by the radiation from
the sQGP: $M=2 M_\mu -0.6$~GeV and $M>1$~GeV.  For more details we
refer the reader to Ref. \cite{Linnyk:2011hz}.

\subsection{RHIC energies}

\begin{figure}
\centerline{    \includegraphics[width=0.43\textwidth]{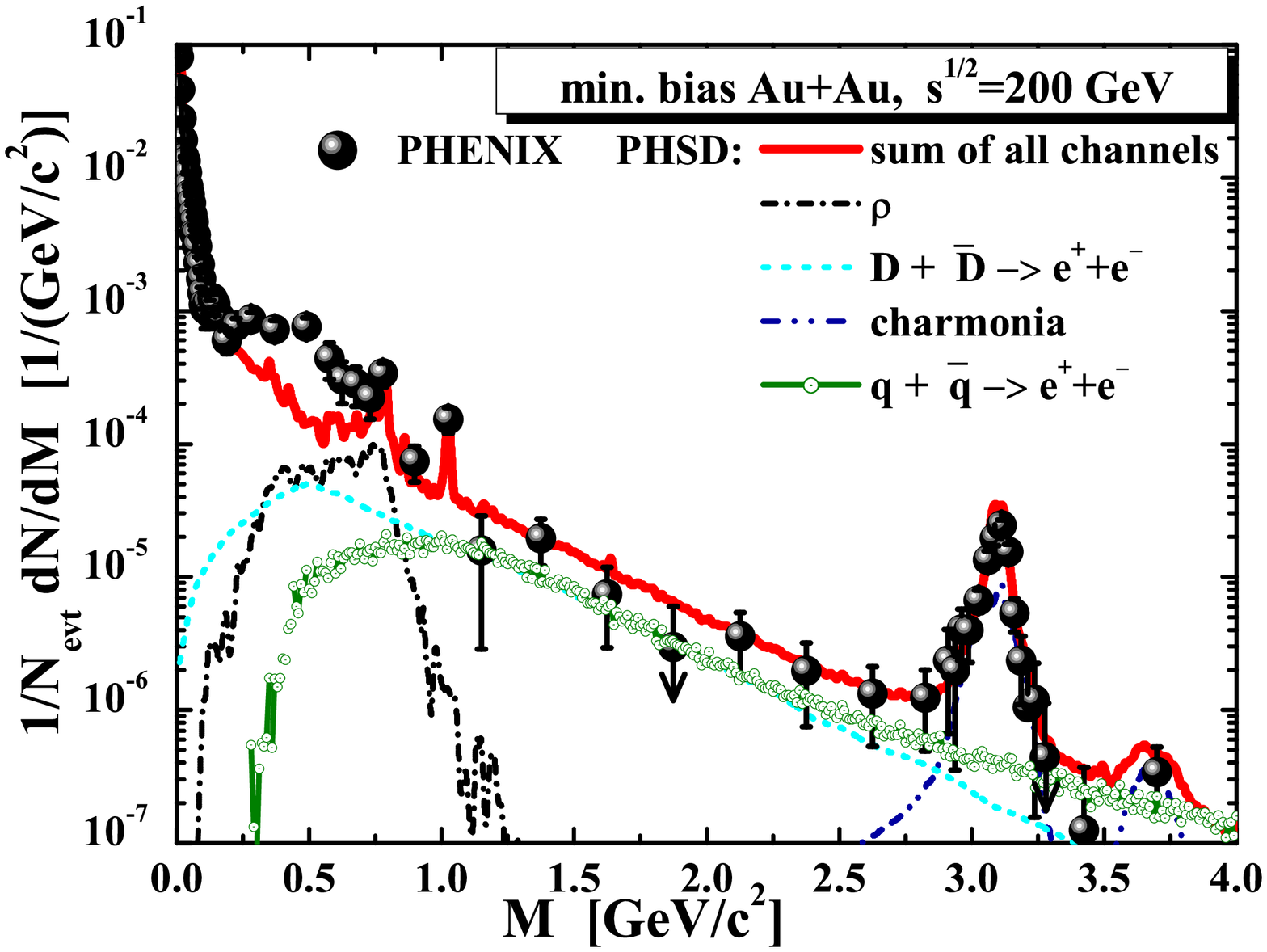}
\phantom{a}\vspace*{-3mm}
\includegraphics[width=0.37\textwidth]{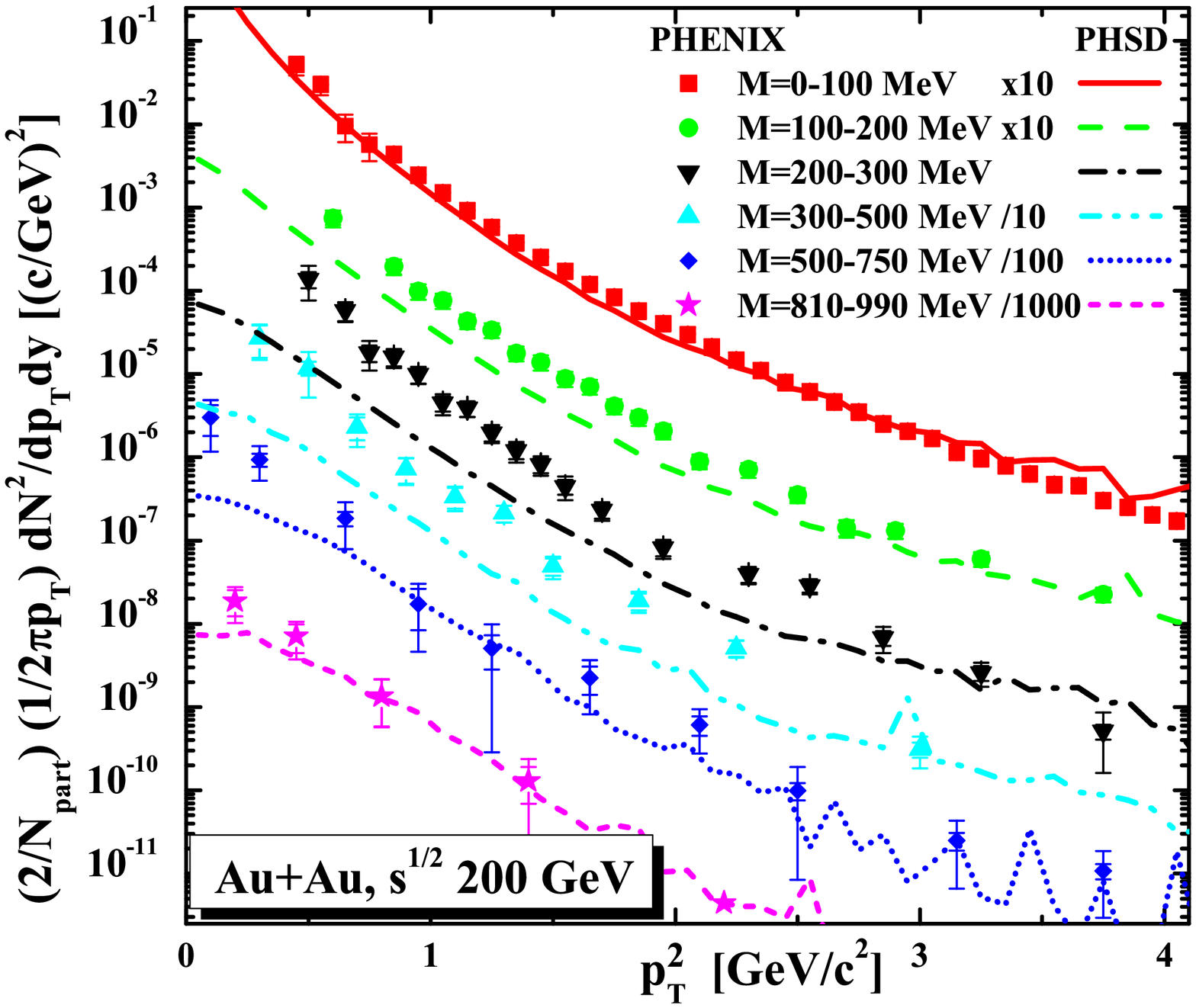} }
 \caption{Left: The PHSD results for the invariant mass spectra of
inclusive dilepton{s} in Au+Au collisions at $\sqrt{s{_{NN}}}$ = 200
GeV within the PHENIX acceptance cuts
in comparison to the data from the PHENIX
Collaboration~\protect{\cite{PHENIX,PHENIXlast}}
The different lines indicate
the contributions from different channels as specified in the
figure.
Right: The PHSD results for the transverse-momentum spectra of
dileptons from minimum bias Au+Au collisions at $\sqrt{s{_{NN}}}$ =
200~GeV in different mass bins compared to the data from {the}
PHENIX {Collaboration}~\protect{\cite{PHENIX,PHENIXlast}}. }
\label{MRHIC}
\end{figure}

Now we are coming to the top RHIC energy of $\sqrt{s{_{NN}}}$ = 200
GeV and present the most important findings from the
PHSD study in Ref. \cite{Linnyk:2011vx}.
In the left part of Fig.~\ref{MRHIC} we show our results for
the invariant mass spectra of inclusive {dileptons in} Au+Au
collisions for the acceptance cuts on
single electron transverse momenta $p_{eT}$, pseudorapidities
$\eta_e$, azimutal angle $\phi_e$, and dilepton pair rapidity $y$:
$ p_{eT}>0.2 \mbox{ GeV}, \  |\eta_e|<0.35, \
-3\pi/16<\phi_e<5\pi/16, \ 11\pi/16 <\phi_e<19\pi/16,\
|y|<0.35.$

In the low mass region $M=0-1.2$~GeV, the dilepton yield in the PHSD is
dominated by hadronic sources and essentially coincides with the
earlier HSD result~\cite{Bratkovskaya:2008bf}.
Note that the collisional
broadening scenario for the modification of the $\rho$-meson was
used in the calculations presented in Fig.~\ref{MRHIC} that underestimates
the PHENIX data from 0.2 to 0.7 GeV substantially.
In contrast, the partonic radiation as well as the yield  from
correlated $D$-meson decays are dominant in the mass region
$M=1-3$~GeV as seen in Fig.~\ref{MRHIC} (left), i.e. in the mass region
between the $\phi$ and $J/\Psi$ peaks. The dileptons generated by
the quark-antiquark annihilation in the sQGP constitute about half
of the observed yield in this intermediate{-}mass range. For
$M>2.5$~GeV the partonic yield even dominates over the D-meson
contribution. Thus, {the inclusion of the} partonic radiation in
{the} PHSD fills up the gap between the hadronic model
results~\cite{Bratkovskaya:2008bf,Manninen:2010yf} and the data of the
PHENIX Collaboration for $M>1$~GeV, however, the early expectation of a
partonic signal in the low mass dilepton spectrum is not verified by
the microscopic PHSD calculations.

In order to investigate the momentum dependance of the "missing
low mass yield", we have calculated the $p_T$-spectra of dileptons in
different bins of invariant mass $M$. In the right part of
Fig.~\ref{MRHIC} we show the measured transverse
momentum spectra of dileptons for minimum bias Au+Au collisions at
$\sqrt{s{_{NN}}}=200$~GeV (symbols) in comparison with the spectra
from {the} PHSD (lines) for six mass bins as indicated in the
figure. Whereas the {PHSD can well describe the} dilepton spectra in
the mass intervals [0,100 MeV] and [810 MeV, 990 MeV]{, it
underestimates the} low $p_T$ dileptons {in} the other mass bins{,
particularly} in the mass bins {[300 MeV, 500 MeV]}.  On the other
hand, high $p_T$ {dileptons} are reproduced quite well by the PHSD
calculations. We conclude that the missing dilepton yield for masses
from 0.15 to 0.6 GeV is essentially due {to} a severe
underestimation of the data at low $p_T$ by up to an order of
magnitude. We recall that at top SPS energies the low $p_T$ dilepton
yield could be attributed to $\pi\pi$ annihilation channels, i.e.
to the soft hadronic reactions in the expansion phase of the system.
{These} channels are{, however,} insufficient to describe the very
low slope of the $p_T$ spectra at the top RHIC energy.

\begin{figure}
    \includegraphics[width=0.47\textwidth]{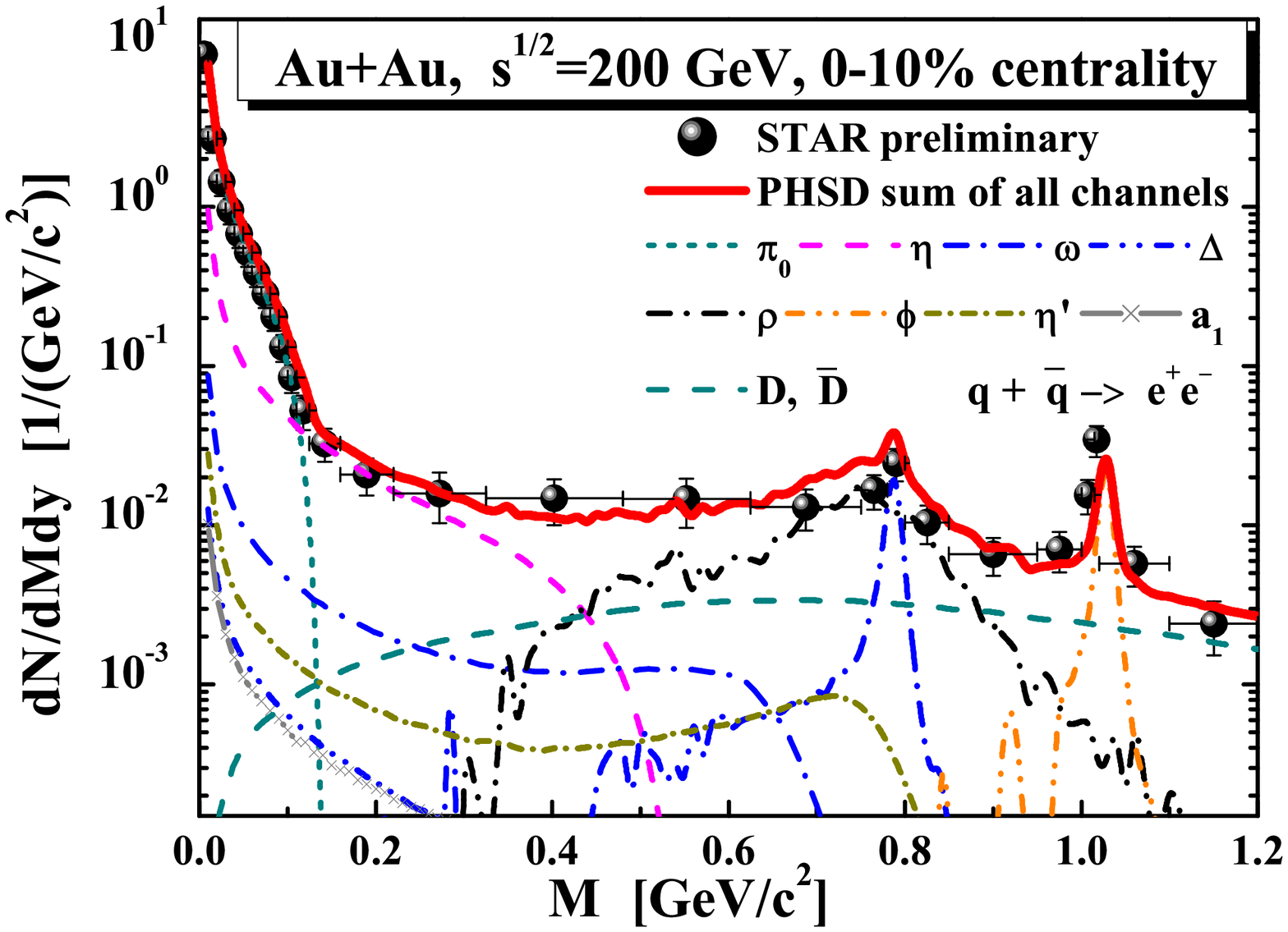}
    \includegraphics[width=0.46\textwidth]{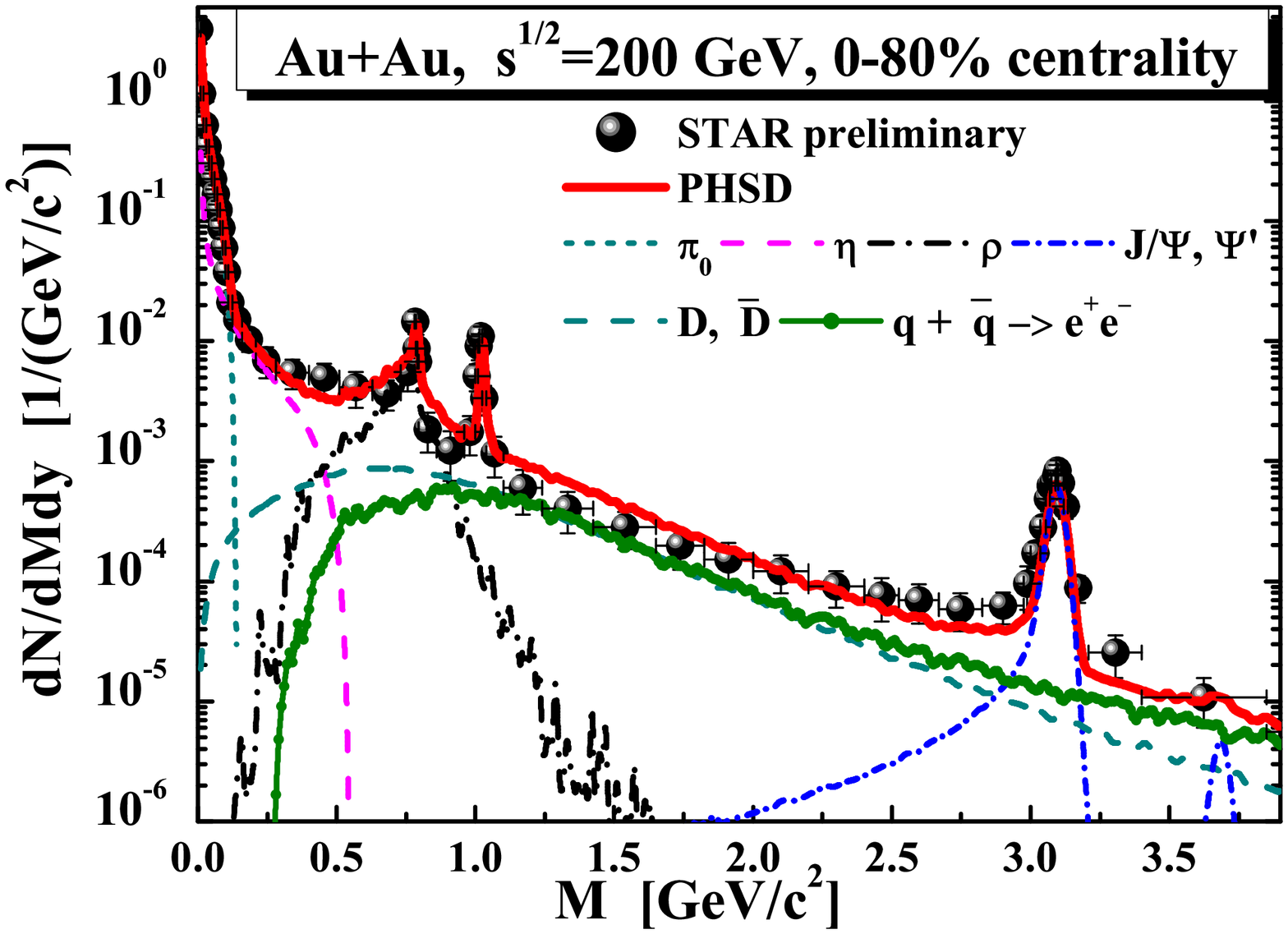}
\caption{The PHSD results for the
{invariant mass} spectra of {dileptons in Au+Au} collisions
at $\sqrt{s{_{NN}}}$ = 200 GeV for $M \! = \! 0 \! - \! 1.2$~GeV (left part)
and for $M\!=\!0\!\,-\,\!4$~GeV (right part)
and 0 - 10 \% centrality within the cuts of the STAR experiment.
The preliminary data from the STAR Collaboration
are adopted from Ref. \protect\cite{Zhao:2011wa}.
} \label{MSTAR}
\end{figure}

In order to shed some light on the 'PHENIX puzzle' we step to a comparison
of the  PHSD predictions with the preliminary STAR data   measured  for
Au+Au collisions at $\sqrt{s{_{NN}}}$ = 200 GeV  with the acceptance
cuts on single electron transverse momenta $p_{eT}$, single electron
pseudorapidities $\eta_e$ and the dilepton pair rapidity $y$, i.e.
$ 0.2<p_{eT}<5 \mbox{ GeV}, \ |\eta_e|<1, \ |y|<1. $
Our predictions for the dilepton yield within these cuts are shown
in Fig.~\ref{MSTAR} for 0-80\% centrality.
One can observe generally a good agreement with the preliminary data from
the STAR Collaboration~\cite{Zhao:2011wa} in
the whole mass regime. Surprisingly, our calculations are also
roughly in line with the low mass dilepton spectrum from STAR in
case of central collisions whereas the PHSD results severely
underestimate the PHENIX data for central collisions.  The observed
dilepton yield from STAR at masses below 1.2 GeV can be accounted for
by the known hadronic sources, i.e. the decays of the $\pi_0$, $\eta$,
$\eta'$, $\omega$, $\rho$, $\phi$ and $a_1$ mesons, of the $\Delta$
particle and the semileptonic decays of the $D$ and $\bar D$ mesons,
where the collisional broadening of the $\rho$ meson is taken into
account.

The discrepancy between the PHENIX and STAR data will have to be investigated closer by the
experimental collaborations.  Furthermore, the upgrade of the PHENIX
experiment with a hadron blind detector should provide
decisive information on the origin of the low mass dileptons produced
in the heavy-ion collisions at $\sqrt{s_{NN}}=200$~GeV.

\section{Conclusions}
We close this contribution by noting that the dileptons are an
interesting probe of the dynamical processes in heavy-ion collisions at
all energy regimes.  The low-mass dileptons provide information on the
vector meson spectral function in the medium whereas the high mass part
from 1.1 to 3 GeV  can  be attributed dominantly to the partonic
annihilation in the QGP phase.  At the higher RHIC and LHC energies the
modifications of the low-mass sector are less pronounced than at SIS
and SPS energies, however, the dilepton emissivity from the sQGP
becomes substantial or even dominant relative to the background from
$D$-meson decays.

\vspace*{1mm}
The authors acknowledge the financial support through the ``HIC for
FAIR" framework of the ``LOEWE" program and the Deutsche
Forschungsgemeinschaft (DFG).

\end{document}